\documentclass[useAMS,usenatbib]{mn2e}
\def\aaps{\ref@jnl{A\&AS}}
\def\mnras{\ref@{MNRAS}}

\usepackage{graphicx}
\usepackage{rotating}
\usepackage{longtable}
\usepackage{lscape}
\usepackage{amssymb,amsmath}
\begin{document}

\def\apj{{ApJ}}                 % Astrophysical Journal
\def\apjl{\ref@jnl{ApJ}}                % Astrophysical Journal, Letters
\def\mnras{{MNRAS}}             % Monthly Notices of the RAS
\def\apjl{{ApJ}}                % Astrophysical Journal, Letters
\def\apjs{l{ApJS}}               % Astrophysical Journal, Supplement
\def\apss{{Ap\&SS}}             % Astrophysics and Space Science
\def\pasp{{PASP}}               % Publications of the ASP
\def\aap{{A\&A}}                % Astronomy and Astrophysics
\def\aapr{{A\&A~Rev.}}          % Astronomy and Astrophysics Reviews

\title[B 0218+357 ]
{On the lensed blazar B0218+357  }

\author[Falomo et al]{
	R. Falomo$^{1}$\thanks{E--mail: {\tt }}
	A. Treves$^{2}$,
	R. Scarpa$^{3,4}$
        S. Paiano$^{1}$,
        and 
        	M. Landoni$^{5}$  \\
      	$^{1}$ INAF -- Osservatorio Astronomico di Padova, Vicolo dell'Osservatorio 5, I-35122 Padova (PD), ITALY \\
	$^{2}$ Universit\`a degli Studi dell'Insubria, Via Valleggio 11 I-22100 Como - ITALY \\
	$^{3}$ Instituto de Astrofisica de Canarias, C/O Via Lactea, s/n E38205 - La Laguna (Tenerife) - SPAIN \\
	$^{4}$ Universidad de La Laguna, Dpto. Astrofísica, E-38206 La Laguna (Tenerife) SPAIN \\
       	$^{5}$ INAF, Osservatorio Astronomico di Brera, Via E. Bianchi 46 I-23807 Merate (LC) - ITALY \\
		}
\maketitle
\label{firstpage}

\begin{abstract}
We  present an optical spectrum  ($\lambda\lambda$ 4000--10500 \AA \ ) of the lensed blazar B0218+357 secured at the 10m GTC and aimed to investigate and clarify the properties of this intriguing system.
We found that the emission line spectrum of the blazar is characterised by only one broad emission line that interpreted as Mg II 2800 $\textrm{\AA}$ yields z~$=$~0.95. In addition we detect narrow absorption lines of Mg II 2800 \AA \ and Ca II (H,K)  and Na I 5892 \AA \  at z=0.68437 $\pm$ 0.00005 due to intervening interstellar gas. No  stellar  absorption features attributable to the lens galaxy are revealed. Thus the assumed redshift of the lens is dubious. 
The continuum spectrum of the object exhibits a remarkable drop towards the short wavelengths likely due to a significant extinction.
This extinction cannot be produced in the lens galaxy at z =0.684 with any value of R$_V$ under the assumption that the intrinsic shape of the blazar is dominated by a power law emission. 
However, the observed continuum is consistent with a power law emission assuming a standard (R$_V$ = 3.1) 
extinction at the source redshift (z~$=$~0.95) as supported also by the presence of Mg II absorptions at the same redshift. HST images of B0218+357 exhibit the double image of the source together with extended image of a face on spiral galaxy. We argue that this galaxy is possibly not  the lensing galaxy but the host galaxy of the blazar.
This has  substantial  consequences on the models of the system  and on the derived values of the Hubble constant.
\end{abstract}


%\end{document}  %%%%%%

\section{Introduction}

B0218+35 is a real wonder of the sky. It is one of the best studied gravitational lenses with a double 
radio image (A+B) separated by 0.33 arcsec, and a clear example of Einstein radio ring , with a diameter similar to the above given separation \citep[e.g.][]{patnaik1993,biggs2001}. A and B exhibit correlated flux variability which is relevant for the measurement of the Hubble constant \citep[e.g.][]{refsdal1964}. 
The redshift of the lens was measured at z=0.684 \citep{browne1993}, while for the source  a redshift z = 0.94 was proposed by \citet{browne1993} and then confirmed by \citet{lawrence1996} . The source is very active at high energies  (e.g. \cite{giommi2012,cheung2014}), and it is classified as a blazar. Till now it is the farthest object detected at  ~$\sim$1 TeV with Cherenkov telescopes \citep{ahnen2016}.

B0218+35 was imaged several times with HST \citep{jackson2000, lehar2000, munoz2004, york2005}.  
The counterparts of the two radio sources are clearly detected, the brighter radio one (A) being the fainter in the optical.  This is generally interpreted as due to absorption in a giant molecular cloud obscuring A, which is supposedly responsible of H I (21-cm)  and other absorption lines detected in the radio band \citep{carilli1993}.  A diffuse emission which is generally associated with the lens galaxy is observed in the HST images. \cite{york2005}    and references therein propose an interpretation as   a spiral viewed face on.

In the course of a systematic study of spectral properties of TeV blazar sources at the GTC 10.4m telescope B0218+35 was observed by our group in 2015 (\cite{paiano2017}).  
In this paper we examine in detail our spectrum and discuss it in the context of the ample literature on the optical properties of the source. The paper is organised as follows: in Section \ref{b0218} we review the optical properties of the system. Section 3 deals with our  spectroscopic observations and analysis. Finally in Section 4 we illustrate and discuss the implications of  our results. In this work we adopt the concordance cosmology with H$_0$ = 70 km s$^{-1}$ Mpc$^{-1}$, $\Omega_m$ = 0.3 and $\Omega_\Lambda$ = 0.7.

\section{Review of optical data of B0218+35} 
\label{b0218}

The first optical spectrum of the source was obtained by \citet{browne1993}
who clearly detected absorption features of Ca II (H, K)  and a very faint absorption attributed to Mg II 2800 \AA  \ of the
lens galaxy at z=0.684. They also propose  the detection of very weak and narrow 
emission lines of [OII] 3727$\textrm{\AA}$ and [OIII] 5007$\textrm{\AA}$.  In addition they suggest the presence of a weak and broad emission line at 5418 \AA \ tentatively identified as  Mg II 2800 \ $\textrm{\AA}$ at z = 0.936 and 
attributed to the source (the blazar). A better spectrum obtained by  \citet{lawrence1996} confirmed the Ca II and Mg II absorptions at z=0.684 and the broad emission at $\sim$ 5400 \AA \ with an associated absorption doublet.
The redshift of the lens galaxy was confirmed through 21 cm HI absorption by \cite{carilli1993} and through molecular gas of CO, HCO and HCN by  \cite{wiklind1995} . 

A superior quality but uncalibrated optical spectrum was obtained by \citet{cohen2003} who confirm the above 
absorption features and clearly detect a strong broad emission line at 5470 $\textrm{\AA}$  identified as Mg II 2800 \ 
$\textrm{\AA}$ yielding a redshift of z~$=$~0.944 for the blazar.  In addition these authors claimed  the detection of emission lines of [OII] 3727, H$_{\beta}$ and [OIII] at z~$=$~0.684 thus attributed to the lens galaxy. Moreover they also suggest the presence   of weak H$_{\beta}$ and [OIII] emission in the red noisy spectrum,  attributed to the blazar at z~$=$~0.944

% Review imaging observations 

\begin{figure}
  \includegraphics[bb = 50 180 500 600, width=0.45\textwidth]{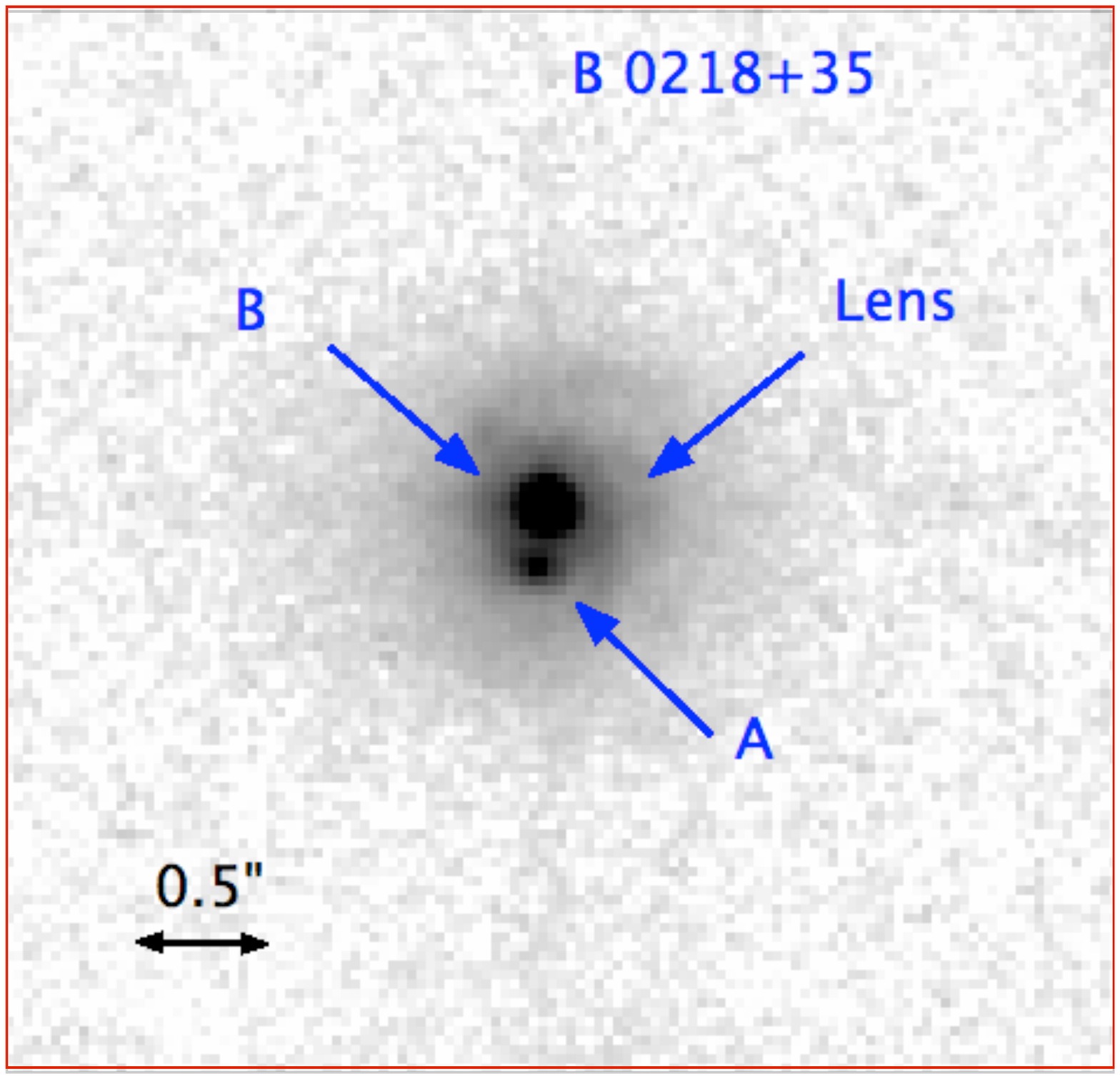}
   \caption{HST ACS WFC image of B0218+35  (filter F814W). The double point-like source is clearly detected surrounded by an extended nebulosity attributed to the lensing galaxy. The separation between the two sources (A and B) is $\sim$ 0.3 arcsec. }
   \label{fighst}
\end{figure}

HST images were obtained by \cite{jackson2000,lehar2000,munoz2004} using WFPC2 and NICMOS and  
by  \cite{york2005} using ACS. From WFPC2 images in V and I filters \cite{jackson2000} clearly detect the two images of the blazar  separated by $\sim$ 0.3 arcsec.   The separation between the two images may be less at optical and infrared wavelengths (318 $\pm$ 5 mas at near-IR \cite{jackson2000} ;  317 $\pm$ 4 \cite{york2005} )  
than that derived at  radio wavelengths ( 334 $\pm$ 1 mas Patnaik et al. 1993, unitis1995).
This  incited some  doubts on the interpretation of B0218+35 as a lensed object. 
 In addition to the double image  HST data show a  "smooth brightness"  around the two sources that is attributed  to 
 the lensing galaxy. The estimated flux density of the lens galaxy is : 0.06, 0.13 and 0.15 
 (10$^{-16}$ erg cm$^{-2}$ s$^{-1}$ \AA$^{-1}$) in the bands F555W, F814W and F160W, respectively \citep{jackson2000}.
 At the redshift of the lens (z= 0.684) these observations correspond to rest frame emission at 
 $\lambda \sim$ 3300, 4800 and 9500 \AA. The F160W observations is close to rest frame I band. Assuming the lens 
 is a spiral galaxy \citep{jackson2000,york2005} the k-corrected absolute magnitude would be M$_I$ (lens galaxy) $\sim$ --23.5 (AB mag).
 Similar results were obtained by \citet{lehar2000}  who, in addition, performed also an image decomposition using the H band  image. It turned out that the lens galaxy is described by an exponential disk with an effective radius R$_e$ = 0.19 $\pm$ 0.01 arcsec.  This would yield  an extremely compact galaxy since at the redshift of the lens it corresponds to only $\sim$ 1.4 kpc. The lens galaxy appears almost centred on the B source (see Figure \ref{fighst}).   
Further HST optical images were secured using WFC of ACS by \cite{york2005}  with the aim of deriving the 
Hubble constant from time delay of flux variations and a mass model of the lens. From these F814W images a spiral structure is clearly apparent. 

%Figure 1 
% Tabella con redshift EW, FWHM
% Confronto z con letteratura 

\section{Spectroscopy and data analysis}

% FIGURE 2 GTC optical spectrum
\begin{figure*}
 \includegraphics[bb= 33 150 550 500, width=0.9\textwidth]{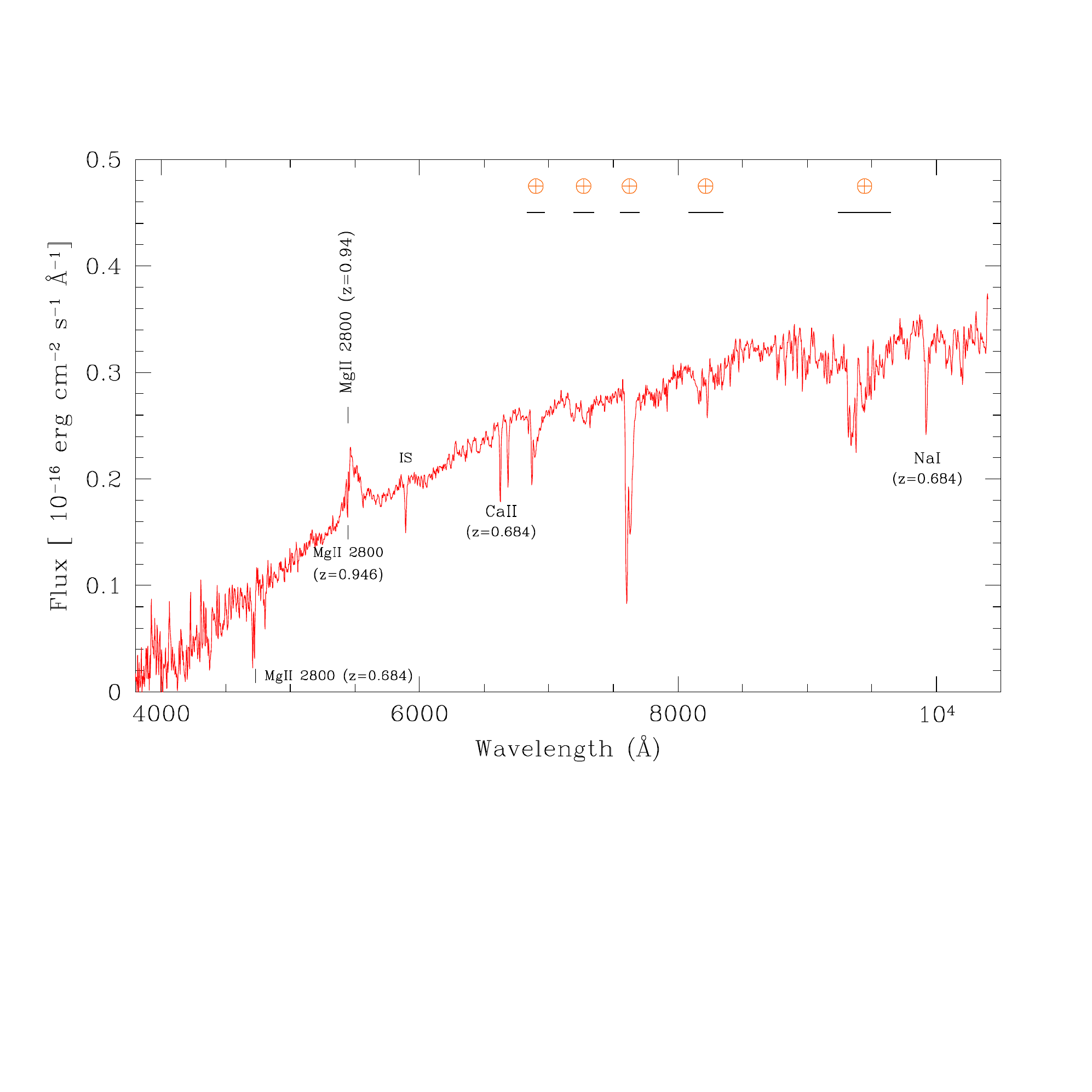}
   \caption{The optical spectrum of B0218+357 obtained at GTC + OSIRIS. The main absorption features are indicated. 
   The spectral regions affected by  telluric absorptions are labelled $\oplus$. IS marks the galactic interstellar absorption of Na I 5892 \ \AA.}
   \label{fig_spec}
\end{figure*}

Observations were obtained on February 2015 
at the GTC using the low resolution spectrograph OSIRIS\footnote{http://www.gtc.iac.es/instruments/osiris/osiris.php}
\citep{cepa2003}. The instrument was configured with the two grisms R1000B and R1000R, in order to cover the spectral range 4000-10000 $\textrm{\AA}$, and with a slit width~$=$~1'' yielding a spectral resolution $\lambda$/$\Delta\lambda$~$=$~800. For each grism three individual exposures were obtained in order to perform optimal cleaning of cosmic rays and of CCD cosmetic defects. 

Data reduction was carried out using 
IRAF\footnote{IRAF (Image Reduction and Analysis Facility) is distributed by the National Optical Astronomy Observatories, which are operated by the Association of Universities for Research in Astronomy, Inc., under
cooperative agreement with the National Science Foundation.}  
and adopting the standard procedures for long slit spectroscopy with bias subtraction, flat fielding, and bad pixel correction. Individual spectra were cleaned of cosmic-ray contamination using the L.A.Cosmic algorithm \citep{lacos}.

Wavelength calibration was performed using the spectra of Hg, Ar, Ne, and Xe lamps providing an accuracy of 0.1~$\textrm{\AA}$ over the whole spectral range.  Spectra were corrected for atmospheric extinction using the mean La Palma site extinction table\footnote{https://www.ing.iac.es/Astronomy/observing/manuals/}. Relative flux calibration was obtained from the observations of spectro-photometric standard stars  secured during the same nights of
the target observation.  No correction for the telluric absorptions was done. The spectra obtained with the two grisms were merged into a final spectrum covering the whole desired spectral range and  calibrated to have the flux at 6231~$\textrm{\AA}$ equal to the photometry found for the target from a short exposure r band image secured before the spectra. 
Finally the spectrum was de-reddened for galactic extinction \citep{cardelli1989} assuming the E(B-V) = 0.06.

\section{Results}
\label{results}

\subsection{The optical spectrum }
\label{optspec}

% FIGURE 3 optical spectrum zoom abs lines 
\begin{figure*}
  \includegraphics[bb= 33 390 500 600,  width=0.9\textwidth]{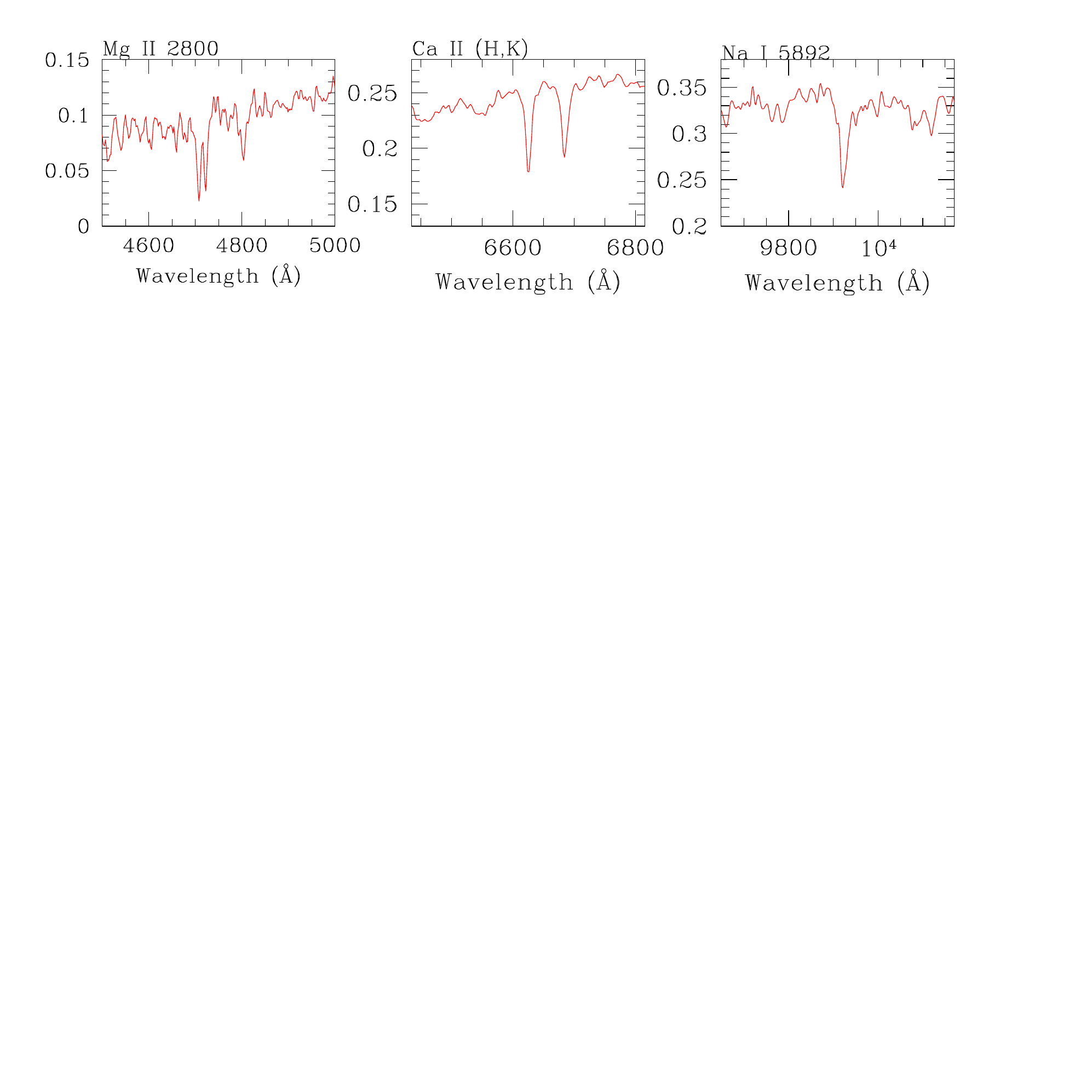}
   \caption{Enlargement of Figure \ref{fig_spec}  showing the  main absorption features in the optical spectrum of B0218+357.    Left: MgII 2800, Middle : CaII, Right Na I at the redshift z = 0.684. }. 
   \label{fig_specz}
\end{figure*}

\begin{figure}
\includegraphics[ width=0.5\textwidth]{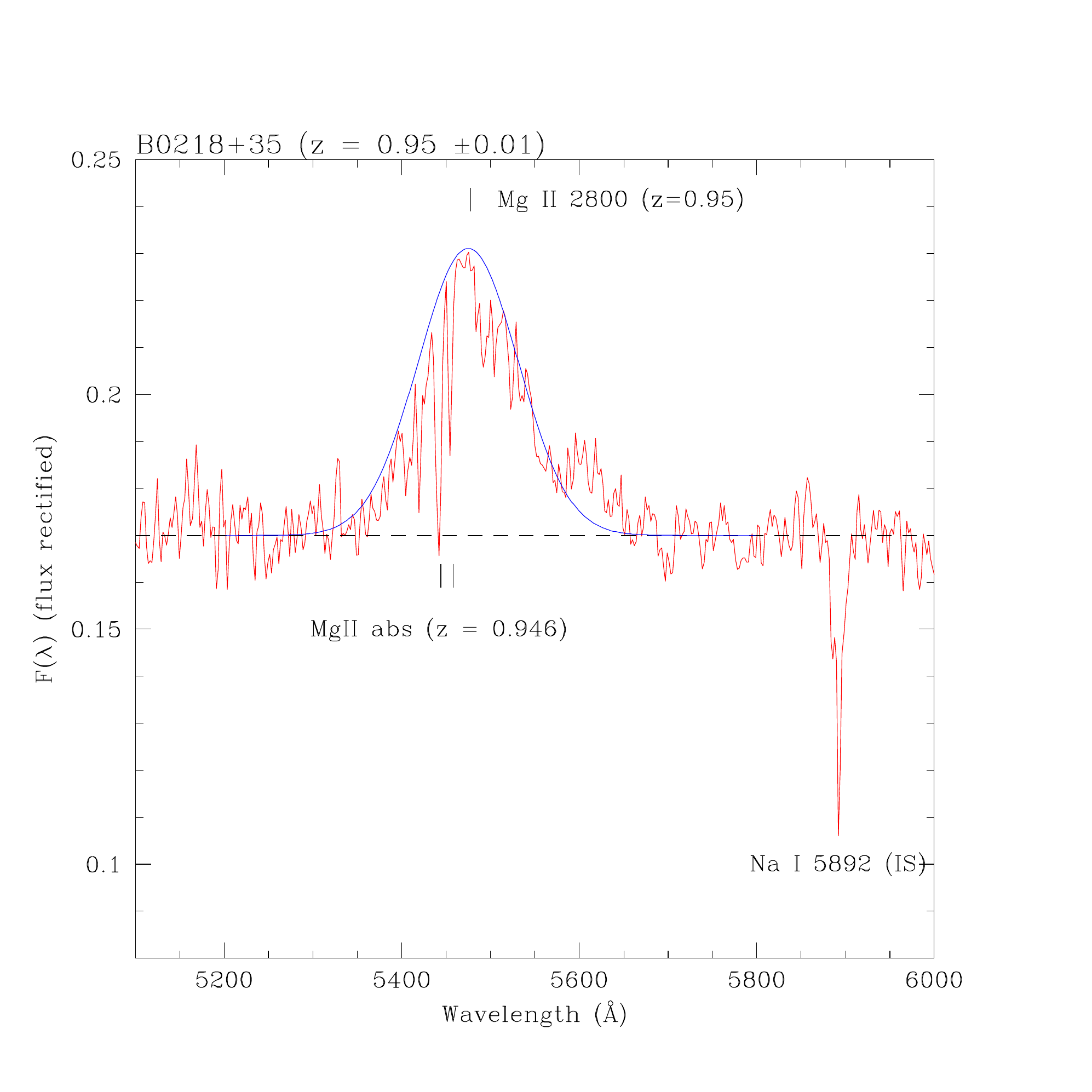}
 \caption{GTC  optical spectrum of B0218+357 in the spectral region of the   Mg II 2800 broad emission at z = 0.95 $\pm$ 0.01. The continuum of the spectrum has been modified to flat shape  to better enhance  the  doublet  of Mg II 2800~\AA  \ absorption features at z = 0.946. }
 \label{fig_specmgii}
\end{figure}

Our final  optical spectrum ( see Figure \ref{fig_spec} ) covers the range between 4000 \AA \ to 10500 $\textrm{\AA}$ with a SNR in the range 30-50.  
The shape of the continuum exhibits a marked decline towards the blue region that is rather unusual for this type of sources and suggestive of a heavy extinction. We confirm the detection of Mg II and Ca II absorption lines at z~$=$~0.684, and in addition we clearly detect an absorption line at 9920 $\textrm{\AA}$ that is identified as Na I 5892 $\textrm{\AA}$ at the redshift of the lens (see Figure \ref{fig_spec} and \ref{fig_specz} and Table 1). 
Contrary to previous spectroscopic results we do not detect the emission lines [OII], H$_{\beta}$ and [OIII] 
reported by \citet{browne1993} and \citet{cohen2003}. 
We note that these features are barely detected by these authors and some of them  occur inside the telluric absorptions of the O$_{2}$ and H$_{2}$O.  
We clearly detect the strong broad emission line at 5480 $\textrm{\AA}$ (EW=35 $\textrm{\AA}$, FWHM=4700 km/s) that, assuming it is attributed to Mg II 2800$\textrm{\AA}$, yields the redshift of z~$=$~0.95 for the blazar.  
On the blue side of this broad emission line (see Figure \ref{fig_specmgii} ) there is an absorption doublet of MgII 
(see Table 1) that is associated with the blazar ($\Delta$ V $\sim$ 1000 km/s).
We stress that in our spectrum we do not detect the  emission lines of  H$_{\beta}$ and [OIII] at z~$=$~0.95 attributed to the blazar by \citet{cohen2003}. These features are placed  in a spectral region heavily contaminated by strong H$_{2}$O atmospheric absorption.  Therefore we conclude that the redshift is based only on one broad emission line identified  with  Mg II 2800$\textrm{\AA}$. 

% FIGURE 5 optical spectrum zoom of MgII 2800.

\subsection{Photometry of B0218+35}

\begin{figure}
\includegraphics[bb = 30 180 500 530, width=0.45\textwidth]{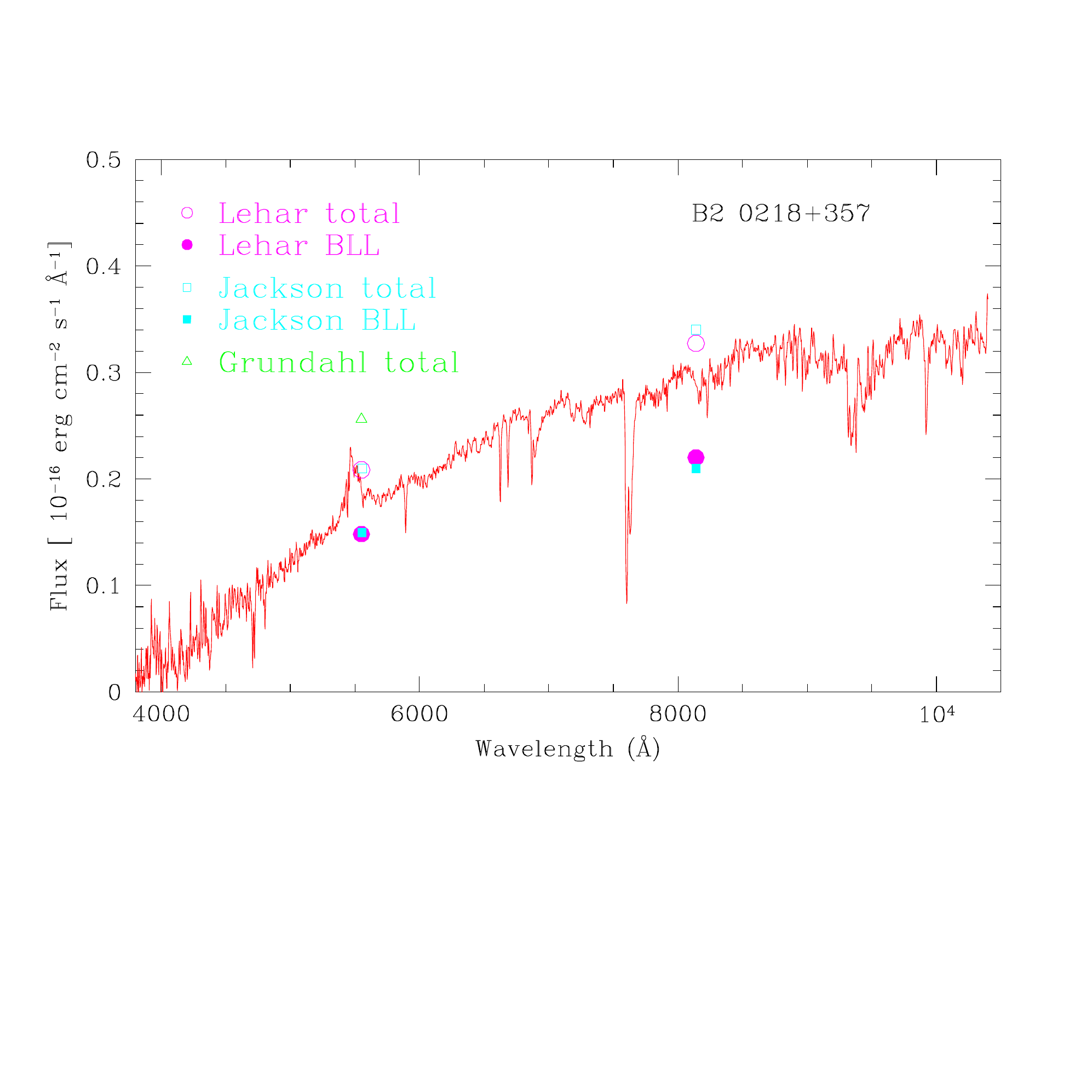}
\caption{Optical spectrum of B0218+35 obtained at GTC + OSIRIS compared with the HST photometry \citep{jackson2000,lehar2000} of the two lensed objects  (filled circles) and  the total flux (blazar plus lens; open circles). The open triangle represent the aperture photometry of the total light by \citet{grundahl1995} }
 \label{sp0218phot}
\end{figure}

In Figure \ref{sp0218phot} our spectrum is compared to  the photometric points derived from the HST images. 
It is noticeable the very small flux variability of the A and B  sources in the HST photometries  covering 1995-2000. 
Comparing the A+B fluxes + galaxy, the photometric agreement is rather good also with our spectrum 
(Figure  \ref{sp0218phot}  ). Other photometry  (e. g.  \cite{grundahl1995},  \cite{ahnen2016}, although sparse indicate a modest optical variability of this source as compared to that of other blazars (e.g.  \citet{falomo2014} and references therein  ).

% Figure 4
\bigskip

\subsection{Continuum shape and reddening}
\label{glens}

Blazars emission in the UV-optical band is characterised by a strong non thermal continuum described by a power law 
(F$_\nu \propto \nu^{-\alpha}$ ) and  a thermal component that can be apparent in the UV region (see e.g. \citet{falomo2014} and references therein). The typical values for $\alpha$ are in the range 0.5-1.5 thus the spectrum is expected to increase towards the blue region (in F$_\lambda$ units) . The substantial drop of the optical  spectrum towards the  blue clearly indicates a strong extinction. Since the colours of the two lensed images are not the same a different extinction for the two light paths was considered. 

The subject has been discussed in detail by \citet{falco1999, munoz2004} who propose a different extinction for A and B and a significant larger value of R$_V$ (R$_V \sim $  7 and  R$_V \sim $  12) with respect to the galactic one  in order to explain the colour difference of the two lensed images in the various bands. The origin of such heavy extinction has been attributed to the lensing galaxy at z =0.684.

From  the photometry of HST images (see Sect. 2)  the diffuse emission due to the lens galaxy represents 
a significant fraction of the total observed  flux at $\lambda \sim$ 8000 \AA \ . Since our spectrum was obtained with a slit width larger than the diffuse emission we assume that the whole  flux from the source (distant blazar and lens galaxy) was gathered. In Figure \ref{sp0218_lens05a} we compare our optical spectrum with a template spectrum of a late type  spiral galaxy (NGC 1057 excluding the prominent emission lines) redshifted at 0.684 and scaled to the flux obtained in the  F814W  band. 

We attempted to fit the observed optical spectrum adding to the galaxy template spectrum the emission of the blazar assuming an intrinsic  power law (F$_\nu \propto \nu^{\alpha}$ ; $\alpha$ = -1) 
extincted by the absorption in the lens galaxy and assuming both a standard extinction law  (R$_V$ =3.1) and 
higher  values of R$_V$ as proposed by \citet{falco1999,munoz2004}.   The fit is optimised by changing the value of E$_{B-V}$ and it is normalised to the observed spectrum at $\lambda =$  7000~\ \AA. \ 
It turns out that while the spectral region at $\lambda >$ 6500~\ \AA \ can be reproduced adequately, at shorter wavelength a significant excess of emission is found (see Figure \ref{sp0218_lens05a} ). 
In order to compensate for this excess a larger value of E$_{B-V}$ could be invoked but in that case the 
region at  $\lambda ~>~$ 6500~\ \AA \ cannot be satisfactorily reproduced. 
We also tried to perform the above fit using different values (-0.5 $< \alpha <$-1.5) 
for the spectral index of the power law emission and the results do not change significantly.
The adoption of a different extinction law as that derived for LMC does not help to better reproduce the observations since the main effect is to reduce or avoid the 2175 \AA \ feature in the extinction curve (see e.g. \citet{cardelli1989}).
Under the above assumptions therefore it is not possible to interpret the observed optical spectrum 
as the combination of a reddened power law and the starlight from a  spiral galaxy template.

\subsection{Lens galaxy features}

\begin{figure*}
 \includegraphics[bb = 30 180 480 530, width=0.9\textwidth]{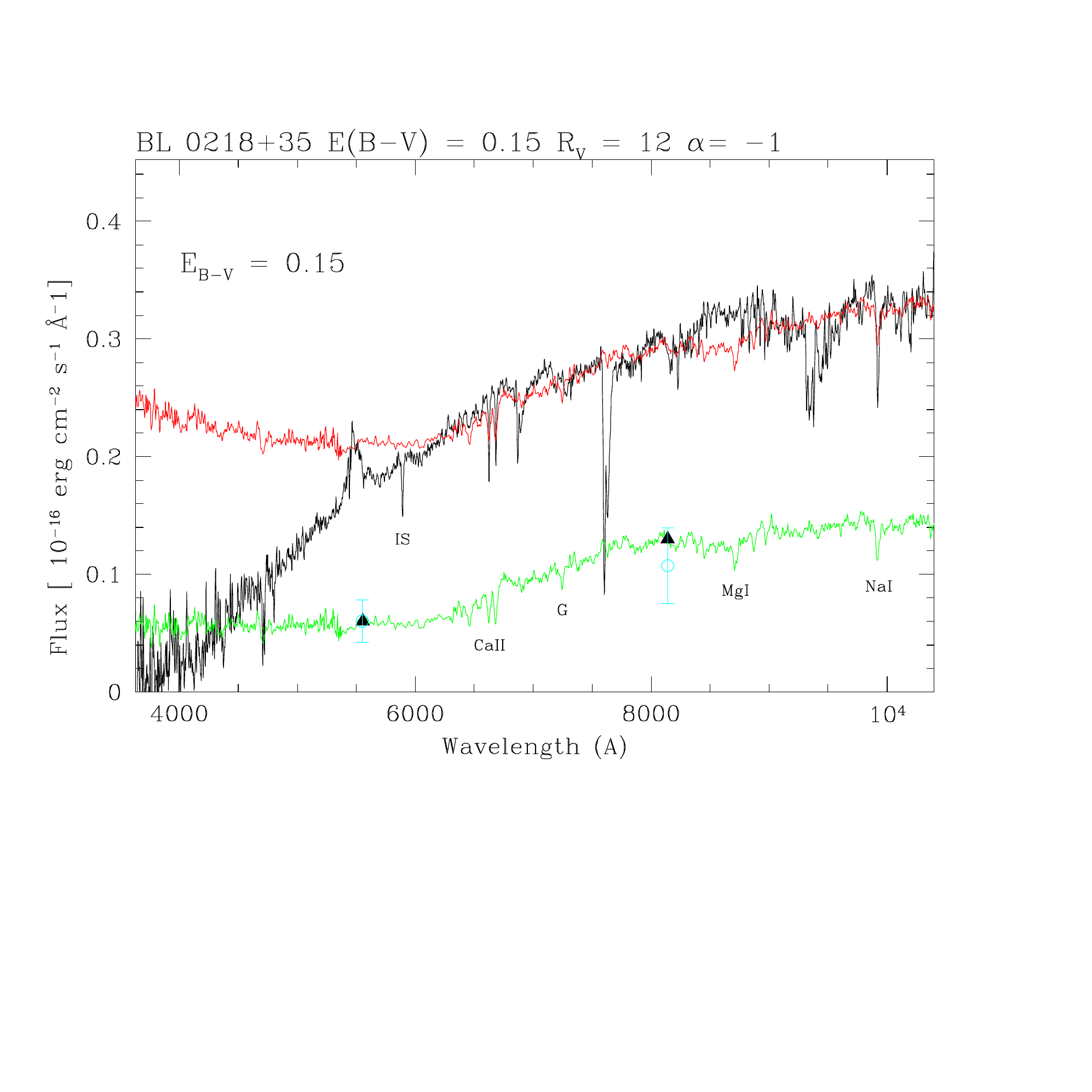}
   \caption{The observed optical spectrum of B0218+35 (black line) compared to a model (red line) obtained adding 
the spectrum of  late type  spiral galaxy  plus  a power law emission (F$_\nu \propto \nu^{-1}$ )   reddened using an extinction law with E$_{B-V}$ = 0.15 assuming R$_V$ = 12 as in   \citet{munoz2004}. The spectrum of the template galaxy ( NGC 1057 with emission lines removed; green line)    has been normalised to the flux given by \citet{jackson2000}    (filled triangles) and by \citet{lehar2000} (open circles).  The fit is remarkably different from the observations in the blue region (see text for details).  } 
   \label{sp0218_lens05a}
\end{figure*}

From the point of view of the spectral features attributed to the lens galaxy we already noted in Sect. \ref{results} 
that we do not confirm the detection of emission lines [OII], H$_{\beta}$ and [OIII] attributed 
\citep{cohen2003} to the lens galaxy. The only confirmed features are thus the absorptions of Mg II and Ca II (see Table \ref{tlines}). 
In addition we also detect Na I absorption at the same redshift (see Figure \ref{fig_spec}, \ref{fig_specz} 
and Table~1 ). These lines are more likely due to interstellar gas (at z = 0.684) than to the starlight from the lens galaxy. In fact these narrow features do not appear diluted by the non thermal 
continuum and no other photospheric absorptions (as G band or Mg I 5175) are revealed in our spectrum 
(see e.g. Figure \ref{sp0218_lens05a}). 

Note also that  no signature of Ca II break is present. These anomalies 
were also partially noted by \citet{jackson2000}. The observed spectrum is at most marginally consistent with the presence of a disc galaxy at z= 0.684 assuming the fluxes derived by HST images (see \citet{jackson2000,lehar2000,york2005} ). 
The redshift derived from K and H absorption lines is : 0.68445 $\pm$ 0.00005 that is marginally consistent with that (  z=0.68466 $\pm$ 0.00004  ) derived from the 21cm neutral hydrogen transition \citep{carilli1993}.

% Figure 6 optical spectrum compared with galaxy + bll but using R=12  

\subsection{The extinction at z = 0.95}
\label{sect_ext}

In Sect. \ref{glens} we showed that the observed optical spectrum cannot be interpreted as a power law plus lens galaxy at z=0.684 with any reasonable value of extinction( E$_{B-V}$ and R$_V$ ) 
The detection of the MgII absorption lines at z=0.946 (see Sect. \ref{optspec} ) suggests the  presence of gas and dust 
associated with the blazar that could produce significant extinction at the source.

\begin{figure*}
 \includegraphics[bb = 27 180 500 500, width=0.9\textwidth]{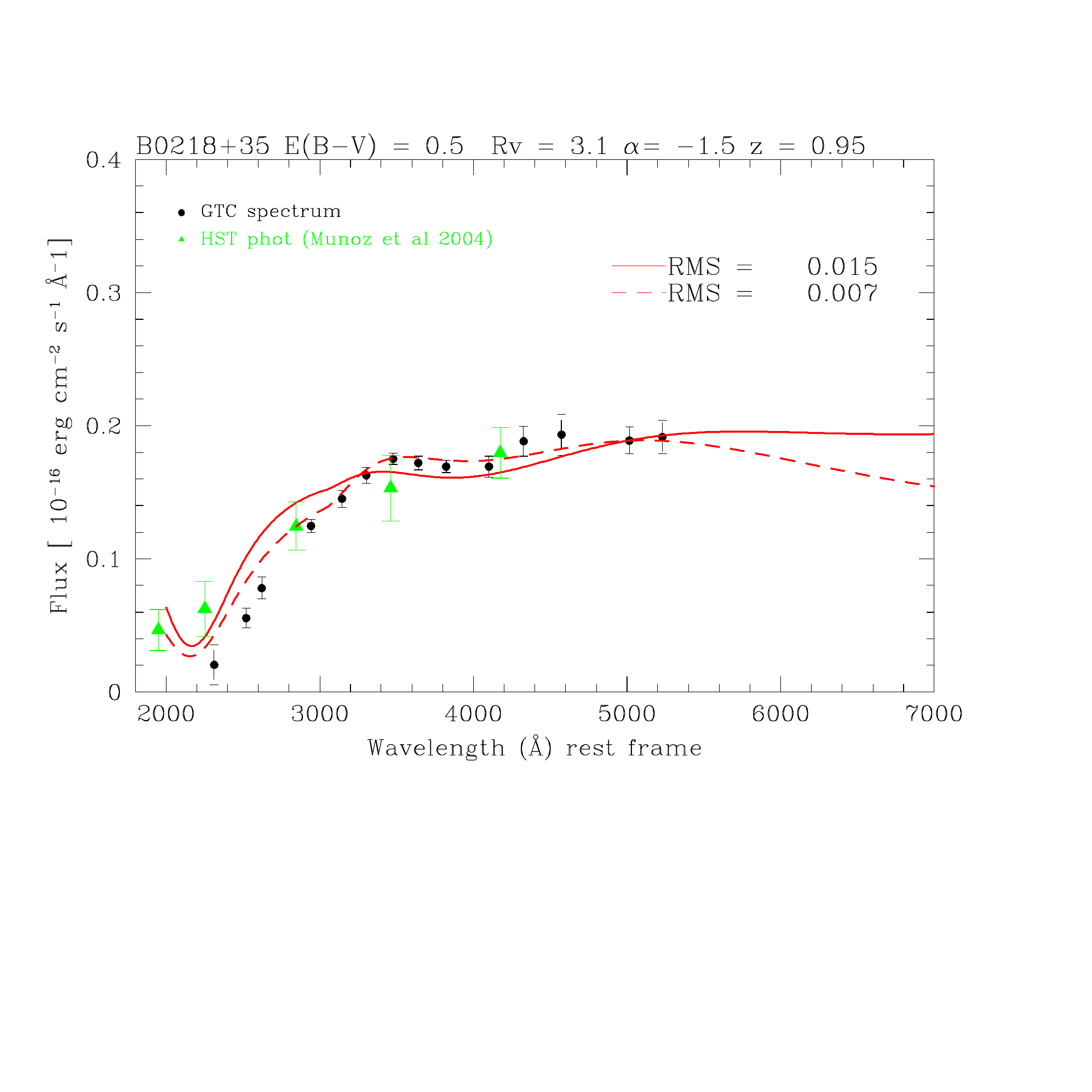}
   \caption{ 
    The continuum (black points) at rest frame of the blazar B 0218+357 (z=0.95) is compared to a model of a power law ($\alpha$ = -1.5)  extincted (red solid line) using R$_V$ = 3.1 and E(B-V) = 0.5.    The optical continuum of the blazar is  derived  subtracting the contribution of the template galaxy to the observed spectrum (see text for details) that are in good agreement with   the HST photometry of the blazar (A+B components;  green filled triangles) obtained by \citet{munoz2004}.   The dashed red line represents the fit above (solid red line) with the addition of an extra extinction at z =0.684 and  R$_V$ = 12 and E(B-V) = 0.1.  } 
 \label{0218ext09}
\end{figure*}

In order to reconcile the observed spectrum with a typical non thermal power law emission of the blazar   we attempted to fit the data assuming that the main extinction occurs very close to the blazar (at z =0.95). In Figure \ref{0218ext09} we report our best fit that assumes  $\alpha$ = -1.5 and galactic extinction law with R$_V$ = 3.1. Under these assumptions we find a significant better fit (weighted RMS 0.016 wrt 0.024) with the data with respect to 
the case where the absorption occurs at z =0.684. 

% optical continuum spectrum compared with power law extinct 0.94

\section{Discussion}

\begin{table}
\label{tlines}
\begin{tabular}{lclll}
\multicolumn{5}{c}{Table 1: Measurements of the spectral features$^*$} \\
\hline \\
Wavelength & EW &  Identification    &  z   \\
( \AA ) & (\AA) & & \\
\hline
%&&&&\\
\multicolumn{5}{c}{Blazar  } \\
&&&&\\
5440.8             & (1.3)    & MgII 2795.5 & 0.946 & a       \\
5454.7             & (1.0) & MgII 2802.7 & 0.946    & a      \\
5470               & $\sim$40  & MgII 2800 & 0.95   & e     \\
\hline
\multicolumn{5}{c}{Lens galaxy } \\
&&&& \\
4708.0            & 6   & MgII 2795.5  & 0.684   & a      \\
4721.5            & 5   & MgII 2802.7  & 0.684     & a     \\
6625.8           & 3.5  & CaII  3933.6  & 0.6844  & a   \\
6684.7           & 3.2  & CaII  3968.5  & 0.6844   & a    \\
9924.8           & 6     & Na I  5982     & 0.684    & a  \\
\hline
\end{tabular}
\caption{* Values in parenthesis are uncertain because of the blend with
 the broad emission. Labels in the last column: a = absorption, e = emission.
}
\end{table}

\subsection{The conventional view}

The generally accepted picture for the source  B0218+35 is that it is a gravitational lensed blazar, and that the lens galaxy at z = 0.684 coincides with the nebulosity detected in the  HST images. The key observations to support this interpretation are the presence  of an Einstein ring in the radio ( \cite{biggs2001} ) and  the radio correlated flux variability  \citep{biggs2001}. There are, however,  some difficulties that is worth mentioning here. 

\begin{figure}
 \includegraphics[bb = 30 180 480 530, width=0.4\textwidth]{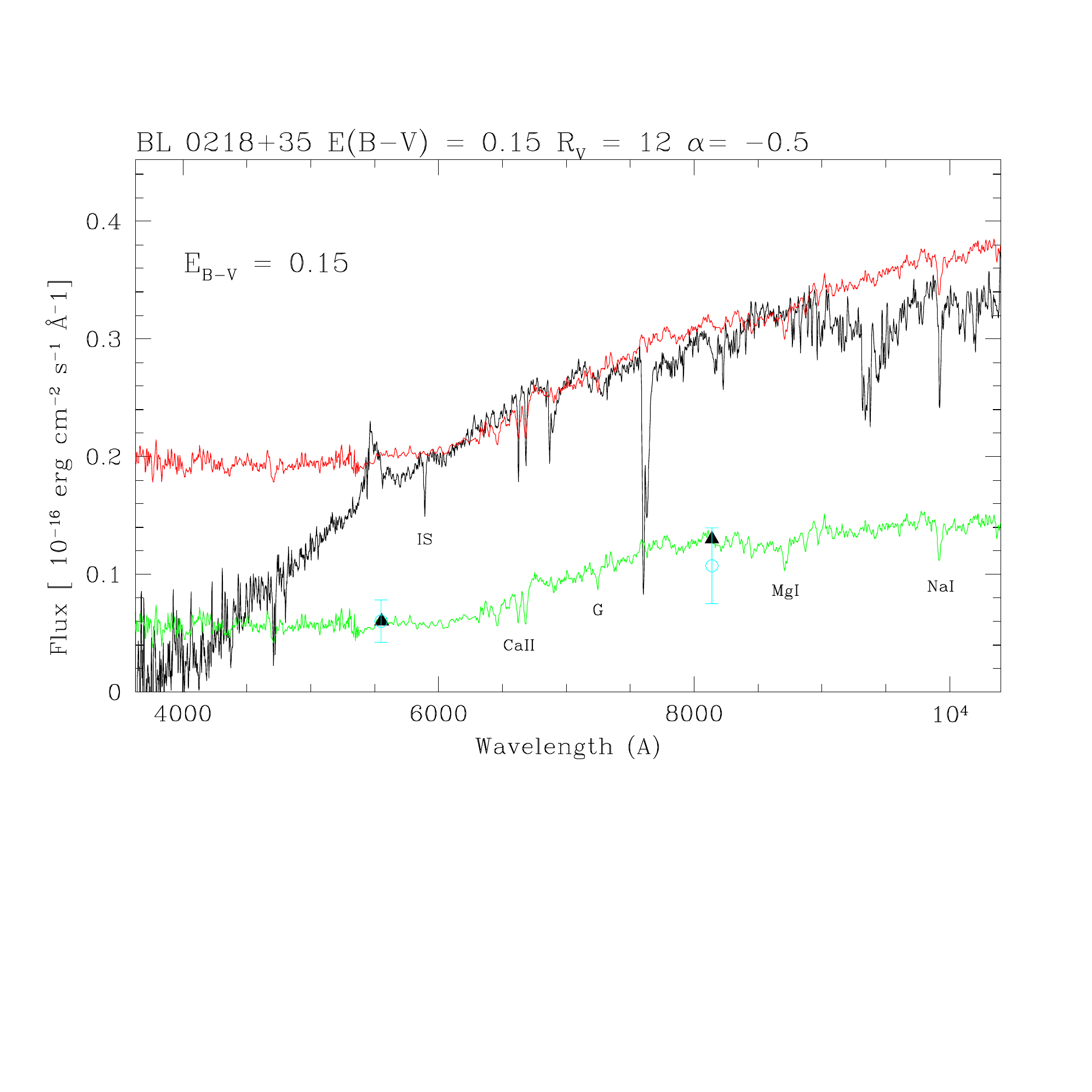} \\
  \includegraphics[bb = 30 180 480 530, width=0.4\textwidth]{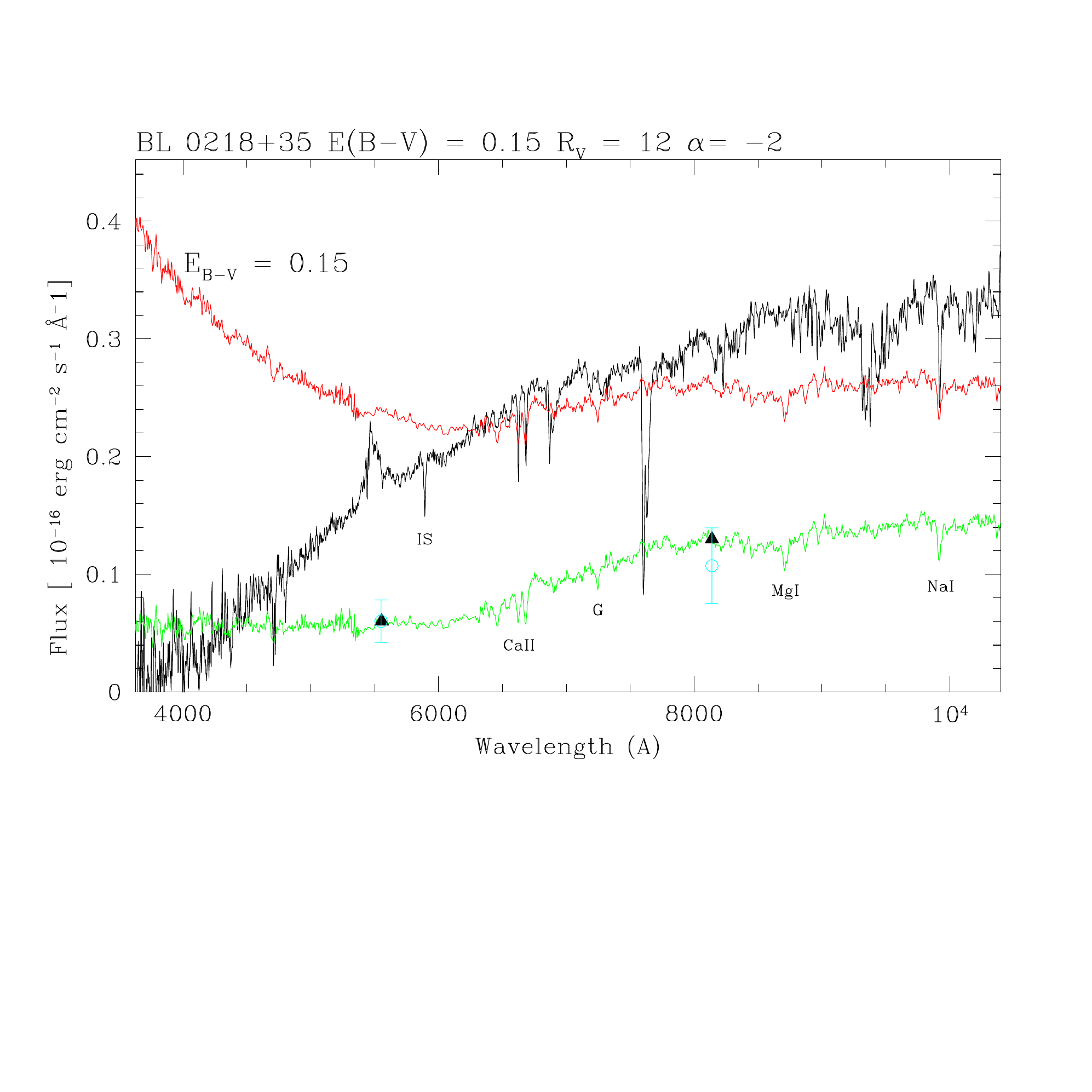}
   \caption{Similar to Figure \ref{sp0218_lens05a} but using different spectral indices for the power law.   } 
   \label{sp0218_lens05a12}
\end{figure}

{\it a) }
We do not confirm the alleged redshift for the lens, showing that all confirmed lines 
are most probably due to intervening absorption gas not necessarily associated to the lens galaxy since no 
photospheric  absorption lines at z= 0.684 are detected. One could argue that for the CaII and NaI lines this is due to the dominance of the gas component, however, one would expect to detect both  MgI 5175  \  \AA \ and G band, which are not observed  (see Figure \ref{fig_spec}).  Although we cannot exclude that the intervening gas belongs to the lens galaxy the observed spectrum appears inconsistent with the superposition of a non thermal continuum plus a  late type galaxy spectrum at the observed fluxes  (see Sect. \ref{glens}) 

{\it b) }
Because of the very small separation ($\sim$ 0.3 arcsec) of the two images of the lensed blazar the mass of the lens is expected to be rather small  \cite[ few times 10$^{10}$ M$\sun$; ][]{grundahl1995, barnacka2016}, in  possible  tension with the  mass derived from the  luminosity (M(R) $\sim$ --23)  of the lens galaxy estimated from HST images \citep{lehar2000,jackson2000}  assuming a very conservative mass to light ratio M/L =1 .  We cannot rule out, however, that only the nuclear region of the galaxy acts as a lens.
 
{\it c) }
The shape of the continuum of the lensed source is inconsistent with a typical power law (F$_\nu \propto \nu^{\alpha}$ )  spectrum of blazars 
\citep[see e.g.][and references therein]{falomo2014}  assuming an extinction law 
 at the redshift of the lens with large R$_V$ as derived from the color difference of the two lensed images.
  In  Figure \ref{sp0218_lens05a} we show the comparison between the observed spectrum and that expected 
 assuming it is the combination of a disc galaxy at z=0.684 and a power law emission ($\alpha$ = --1) taking an extinction with R$_V$ = 12.  A substantial difference is apparent at $\lambda < 6000 \ \AA$ between the fit and the observed spectrum.  The fit becomes even worse if a smaller or a greater value of the spectral index  is used 
 (see Figure \ref{sp0218_lens05a12}). 

\subsection{An alternative scenario}

Because of the above anomalies we are here considering an alternative scenario  of B0218+35. 
Since neither  stellar absorptions nor emission lines are firmly detected at z = 0.684 from the image of the galaxy it is plausible that  the nebulosity detected with HST is instead due to  the host galaxy of the blazar.  Taking z = 0.95, the absolute magnitude then becomes  M$_I$  $\sim$ --24.1 (M$_R$ $\sim$ --23.5 ) that is consistent with typical host galaxies of  QSO.  In the optical spectra we detect absorption lines of Mg II 2800 at the redshift  ($\Delta$ V $\sim$ 1000 km/s) of the blazar. This is suggestive of the presence of gas and dust  in the line of sight that is associated to the host galaxy, not the lens. 
Under these assumptions it is possible to reproduce the observed spectrum with an intrinsic  power low emission that is 
reddened by a standard (R$_V$ = 3.1) galactic extinction law (see also Sect. \ref{sect_ext}  and Figure \ref{0218ext09}).  
A further improvement of the fit is obtained if one adds an extra extinction at z=0.684, E(B-V) =0.1 and R$_V$=12 (see Figure \ref{S0218_HST_ACS_ima}). In this scenario the non detection of the stellar absorption  lines from the host galaxy could be explained since at the redshift of the source (z=0.95) all prominent ones fall in a region highly affected by the atmospheric absorption bands (e.g. Ca II would be at 7730-40 \AA). 

The consequences of this picture are :

{\it a)}  The spectral absorption lines  at z=0.684 are due to the halo gas of an intervening  galaxy.   There are various examples of absorption features in BL Lac objects, that are due to this effect \citep{landoni2014, paiano2017}, 
and not always the identification of the galaxy is obvious. 

{\it b)}
 Apart possibly for  its redshift, the properties of the lens galaxy (luminosity and size) remain unknown. The strong absorptions at z=0.684 together with the lack of stellar lines suggest it is a low luminosity gas dominated galaxy, as expected from the small separation of the two gravitationally lensed images. This might imply a substantial revision of the lens models for B0218+357. 

{\it c)} 
Under this scenario one would expect that  also the image of the host galaxy of the blazar be distorted by the lensing effects. However, since the mass of the lens is likely very small it is not surprising that the distortion be not detectable by the HST images. Moreover the very tiny difference of position ($\sim$ 50 mas) between the centre of the extended nebulosity  and of image B, if real, could also be consisted with  a low mass lensing galaxy.

\section{Summary and Conclusions}

We presented a new optical ($\lambda\lambda$ 4000--10500 \AA \ ) spectrum of the  lensed blazar B0218+357  that was recently detected at TeV energies confirming the blazar nature of this source. The analysis of this optical spectrum reveals a number of  features  that allow us to elucidate some of the puzzling properties of this source and to emphasize peculiarities that suggest possible different interpretation of the system.

%  HST images 
\begin{figure}
 \includegraphics[width=0.45\textwidth]{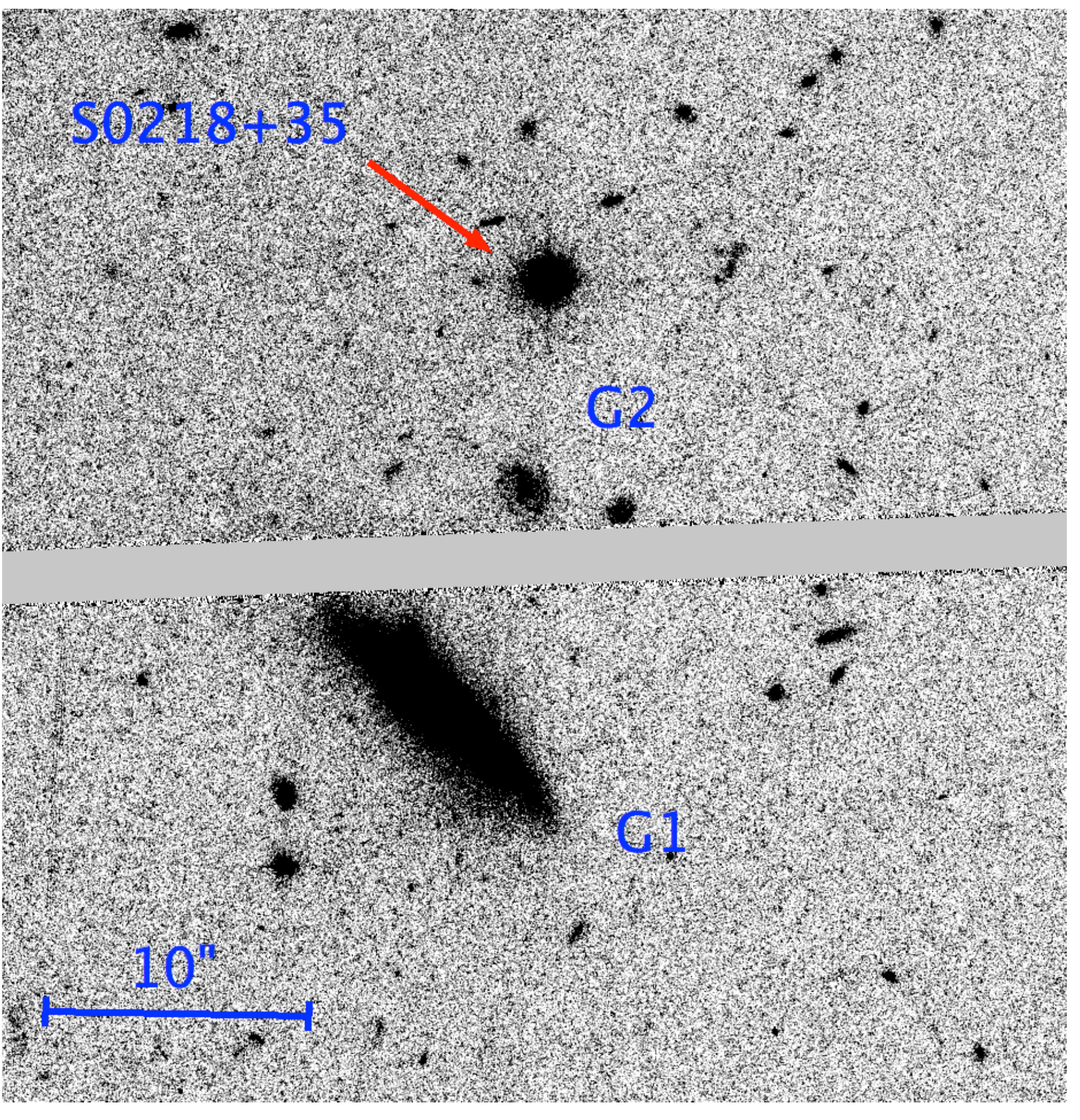}
   \caption{ HST ACS image of the field around S0218+35. The target is close to the separation between the two 
   detectors of the WFC. Two nearby galaxies are observed: G1 at 19 arcsec (PD = 130 kpc) and G2 at 9 arcsec (PD = 65 kpc) at z = 0.684. See text for more details.  } 
\label{S0218_HST_ACS_ima}
\end{figure}

From the analysis of our GTC spectrum we confirm the presence of a broad emission line of Mg II 2800 $\textrm{\AA}$ yielding a redshift of z~$=$~0.95. However, we contradict the detection of other emission lines claimed by \citet{cohen2003} thus, although likely the redshift of the blazar is based only on one broad emission feature. 
On the other hand we confirm the detection of absorption features of Mg II 2800 \AA \ and Ca II (H,K) at the same z   and furthermore we detect also Na I 5892 \AA \ absorption at the same redshift. 
We argue that these lines are arising from interstellar gas likely associated to an intervening galaxy 
since no other absorption features arising from the stellar population (i.e. G band, MgI 5175 \AA \ ) of the lens galaxy are found.

In the field around the target (see Figure \ref{S0218_HST_ACS_ima} ) there are two galaxies. One (G1) is relatively bright (I $\sim$ 17.8) at 19 arcsec  S from the target and another much fainter (I $\sim$ 21.8) at 9 arcsec S. 
While the former is unlikely to be at the redshift of the intervening absorptions the latter is a plausible candidate for the associated intervening absorption. In fact if the galaxy G2 be at z =0.684 then the projected distance from B0218+357 would be only $\sim$ 65 kpc and its absolute magnitude M$_I$ = -- 21.8. 

It is apparent that the nature of B0218+357 is still  rather puzzling. In order to fully  understand the properties of this system spectroscopy of the nebulosity and of the individual  A and B sources
 together with spectra of the galaxies in the immediate environments  are mandatory. 
 The detection of stellar spectral lines from the nebulosity would yield unambiguously its redshift.
These observations require high sensitivity and adequate angular resolution and are likely achieved  in the near future with the  new generation of telescopes and  instrumentations as JWST, TMT and E-ELT.

\bibliographystyle{aasjournal}
\bibliography{bllacs_biblio}

\end{document}